\documentstyle[twocolumn,aps]{revtex}

\newcommand{\beq}{\begin{equation}}
\newcommand{\eeq}{\end{equation}}
\newcommand{\beqa}{\begin{eqnarray}}
\newcommand{\eeqa}{\end{eqnarray}}
\newcommand{\beqar}{\begin{eqnarray*}}
\newcommand{\eeqar}{\end{eqnarray*}}

\def \la {\langle}
\def \ra {\rangle}
\def \up {\uparrow}
\def \down {\downarrow}
\input epsf

\begin{document}

\title{Remote generalized measurements (POVMs) require non-maximal
entanglement}

\author{ {\large Benni Reznik} \\
 { School of Physics and Astronomy, Tel Aviv
                      University, Tel Aviv 69978, Israel.}   }
\date{Mar. 13 2002}

\maketitle

\begin{abstract}
{We show that non-maximal entangled states can be used
for implementing, with unit probability,
remote generalized measurements (POVMs).
We show how any n-qubit POVM can be applied remotely and
derive its entanglement cost.
The later turns out to be equal to the entanglement capability
for a class of POVMs.
This suggests a one-to-one relation between sub-sets of POVM
operations and entanglement.}
\end{abstract}
\vspace{.4cm}

Although quantum entanglement has been a major research topic
for the last decades,
the nature of the relation between entanglement non-locality and
the properties of physical interactions
is a fairly new subject.
During the recent years it was realized that entanglement can
be used as a resource for implementing various types
of remote interactions and operations [1-11].
Optimal ways for generating entanglement using a given
interaction have been searched [12-14].
In particularly in ref. \cite{popescu} an
interesting qualitative general conjecture has been raised,
while  \cite{cirac1,cirac2} pointed out  a detailed connection
between entanglement and operations.
Nevertheless, the fundamental question mentioned above
still seems to be open. In this letter we address this question.

Most known implementations of non-local operations,
either require maximal-entangled states,  or become
probabilistic when non-maximal states are used.
Nevertheless in exceptional cases
non-local operations can be performed
with unit probability and non-maximal entanglement
\cite{cirac1,cirac2}.

The main purpose of this letter is to show that
non-maximal entangled states can in-fact be used
for implementing, with unit probability,
a remote generalized measurement, usually referred  to as a POVM (positive
operator
valued measure) \cite{peres}.
We show that any n-qubit POVM can be measured
remotely with certainty by using
local operation and classical communication (LOCC), and
single non-maximal entangled state.
We also provide a general relation between an n-qbit
POVM and the required amount of entanglement, which turns out
to be equal to the entanglement capability of the POVM.
We can hence classify the space of POVM operations
into sub-sets, each having a definite entanglement measure.
This  suggests a one-to-one non-asymptotic relation between
sub-sets of POVM operations and entanglement.

It will be helpful to begin with a concrete example.
Suppose that one bit is encoded by two non-orthogonal states
\beq
|\psi_\pm\ra = \alpha|0\ra \pm \beta |1\ra .
\eeq
This bit cannot be retrieved back with certainty,
however a generalized measurement,
allows us to distinguish (sometimes) with certainty between $|\psi_\pm\ra$.

Suppose we hand Bob the qubit,
and informs $only$  Alice (that is located remotely), what
are the possible states $|\psi_\pm\ra$. How can Alice and Bob
measure the POVM?
Surely, Bob can teleport \cite{teleportation} his state to Alice
which then proceeds to perform the POVM.
We show however that the POVM can
be applied with optimal efficiency with less than one
ebit of entanglement. (i.e. without teleportation).

\begin{figure} \epsfxsize=2.5truein
      \centerline{\epsffile{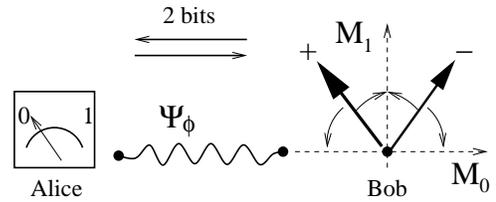}}
\vspace {0.5cm}
  \caption[]{ Alice applies remotely either $M_0$ or
  $M_1$ on Bob's state. In the first case Bob can proceed
  to measure in the x-direction and distinguish with certainty
  between $\psi_\pm$. The operation uses a non-maximal
  entangled state $\Psi_\phi$ and  one bit in
  each direction.}
    \label{prepare} \end{figure}

%plot of the problem

Let us consider first a POVM that allows to
distinguish (sometimes) between $|\psi_\pm\ra$ with certainty.
A generalized measurement can always be described as a unitary
operation acting on the system and an ancilla, followed by
a projective measurement of the ancilla. In the present case
we can use another qubit as the ancilla and operate
on both the unitary $U$ such that
\beq
U|0\ra |\psi_\pm\ra =|0\ra M_0|\psi_\pm\ra +|1\ra M_1|\psi_\pm\ra ,
\label{unitary0}
\eeq
where
\beqa
M_0 &=& {1\over2}(1+{\alpha\over\beta})  + {1\over2}
(1-{\alpha\over\beta})\sigma_z , \nonumber \\
M_1 &=& -{\sqrt{\beta^2-\alpha^2} \over 2\beta}  (1-\sigma_z) ,
\eeqa
are non-unitary operators and $M_0^\dagger M_0 + M_1^\dagger M_1
=1$. $M_0$ and $M_1$ are  usually referred to as Kraus operators
\cite{preskill}.
After measuring the ancilla, we find whether $M_0$ or $M_1$ have been
generated. $M_0$ acts to rotate the non-orthogonal states $|\psi_\pm\ra$
to $|0\ra+|1\ra$ or $|0\ra-|1\ra$, in which case we can distinguish with
certainty
between $|\psi_\pm\ra$ by measuring $\sigma_x$ of the system.
If the ancilla is observed to be in the state $|1\ra$, $M_1$
maps both states into $|1\ra$, and the encoded
information is completely lost.
In this example the POVMs are $F_i = M_i^\dagger M_i$.

Let us now see how  Alice can assist Bob remotely to
perform this POVM (Fig. 1).
While in the usual procedure the ancilla is at the hands of Bob, here
we will have the ancilla  with Alice.
We start with a shared entangled state
\beq
\cos\phi |0\ra_A |0\ra_{b} +\sin\phi |1\ra_A|1\ra_{b},
\eeq
where the angle $\phi$ will be fixed in the sequel.
Bob performs a controlled-NOT
in the $\sigma_z$ direction, with the qubit
$b$ as the control, and $|\psi_\pm\ra_B$ the target.
He next measures $b$ in the $\sigma_x$ direction,
and sends one bit to inform Alice the outcome.
Alice uses this information and by performing
a $\pi$ (or zero) rotation along the $z$-direction
obtains the state
\beq
(\cos\phi |0\ra_A +\sin\phi |1\ra_A\sigma_z) |\psi_\pm\ra_B ,
\eeq
where we ignored the state of $b$ that factors out.

Next Alice applies a rotation that maps
\beqa
|0\ra_A &\to& \cos\phi |0\ra_A - \sin\phi|1\ra \nonumber_A,  \\
|1\ra_A &\to& \sin\phi |0\ra_A + \cos\phi |1\ra_A ,
\eeqa
with the angle $\phi$ determined from:
\beqa
\cos^2\phi  &=&
{1\over2}(1+{\alpha\over \beta} ) ,  \nonumber \\
\sin^2\phi  &=&
{1\over2}(1- {\alpha \over\beta}) .
\eeqa
This transformation maps the entangled state of the ancilla and
Bobs system to the desired form in the right hand side of equation
(\ref{unitary0}).
To completes the process Alice and Bob measure their systems and
Alice sends one bit according to her result to Bob.

The main point is that by using only shared entanglement and
LOCC,
we reached an entangled state
of same form  as in the right hand side of  eq. (\ref{unitary0}).
The success probability, $\la \psi_\pm| F_0|\psi_\pm\ra$,
 is identical to that in an ordinary, local, POVM.

The entanglement consumed in the process
\beq
E_{POVM} =-\cos^2\phi\ln\cos^2\phi -\sin^2\phi\ln\sin^2\phi ,
\eeq
is generally less than  one ebit.
It goes to zero for $\la \psi_+|\psi_-\ra\to 0$,
and tends to $E_{POVM}\to 1$ when  $\la \psi_+|\psi_-\ra\to 1$.
The classical communication cost is two bits one in each
direction.

Clearly, in this example, the POVM dictates the amount of entanglement needed.
Have we used instead a  maximally entangled state,
we could still generate $M_0$ remotely, but with a smaller
probability of success.
We can still make use of a maximally entangled state.
Alice first dilutes to a non-maximal state, which she can do
with unit probability \cite{nielsen},
and then applies the above procedure.
In passing we remark that the one bit sent from Bob to Alice is random.
On the other hand the bit sent from Alice to Bob is biased according to the
success probability of the POVM.
Hence over many trails Bob can gain information on the
inner product $\la \psi_+|\psi_-\ra$.

We next consider the problem in more general terms.
Every POVM on system $B$ may be realized by letting $B$
first interact with an ancillary system $A$ in a standard initial state,
and then observe $A$.
\beq \label{unitary}
U_{AB} |0\ra_A |\psi\ra_B = \sum_\mu |\mu\ra_A M_\mu |\psi\ra_B ,
\eeq
where the Kraus operators, $M_\mu= {\ _A}\la \mu|U_{AB}|0\ra_A$,
satisfy $\sum M_\mu^\dagger M_\mu=1$.
The corresponding POVM,   $F_\mu=M^\dagger_\mu M_\mu$,
appears with the probability distribution
${\rm Prob}(\mu) ={\ _B}\la \psi|F_\mu|\psi\ra_B$.

The measurement of the ancilla $A$ realizes
a particular (non-unitary) transformation on the system.
On the other hand if we measure the
ancilla in a different basis, or equivalently  first apply
a unitary $U_A$ we obtain
\beq
U_A\sum_\mu |\mu\ra_A M_\mu = \sum_{\eta,\mu} |\eta\ra_A
U_{\eta\mu}M_\mu =
\sum_\mu |\mu\ra_A N_\mu ,
\eeq
where
\beq
\label{unitarymap}
N_\mu = \sum_\rho U_{\mu\rho}M_\rho .
\eeq
Therefore, a unitary transformation on the ancilla gives rise to
a new set of Kraus operators.

Before we proceed it is instructive to compare our problem with
the superoperator picture.
If we do not observe the ancilla, the effect of $U_{AB}$
on the subsystem $B$ is described by a superoperator
$ \$_B\rho_B = \sum M_\mu \rho_B M_\mu^\dagger$.
Two unitary related sets, such as $M_\mu$ and $N_\mu$ above,
then represent the same $\$_B$.
Nevertheless, when we do observe the ancilla, as in our case,
we learn which Kraus operator has been realized on the system.
Hence, two unitarily related Kraus sets generally give rise to
inequivalent POVMs.

The realization of the POVM requires an interaction
$U_{AB}$ between the ancilla and the system. Our main goel
is to find how to construct this transformation, using
entanglement and LOCC,
when the ancilla is located remotely with Alice.
To this end we start with some preliminary steps.

\noindent
{\bf Definition 1.}:  A a set of Kraus operators will be
defined as an orthogonal set if
\beq      \label{inner}
(M_\mu,M_\eta)\equiv{1\over N_B}{\rm Trace} M_\mu^\dagger M_\eta =
c_\mu\delta_{\mu\eta}
\eeq
where $N_B$ is the dimension of Bob's system.
For example,  $M_\mu =\alpha_\mu \sigma_\mu$, $\mu=0,...3$,
where $\sigma_k$ $k=1,2,3$ are Pauli matrices and $\sigma_0=1$,
constitute an orthogonal set.

Notice that the unitary transformation (\ref{unitarymap})
generally {\em does not} preserve the inner produce (\ref{inner}).
Clearly, if $N_\eta = \sum_\mu U_{\eta\mu} \alpha_\mu\sigma_\mu$,
${\rm Trace}( N_\mu^\dagger N_\eta) \neq c_\mu \delta_{\mu\eta}$,
unless all $\alpha_\mu$ are equal. Therefore, in certain cases,
by applying  the
unitary transformation (\ref{unitarymap})
we may obtain from a non-orthogonal
an orthogonal one.

\noindent
{\bf Definition 2.}:
A set of Kraus operators $M_\mu$
will be said to admit an orthogonal equivalent set, or shortly OE,
if it is unitarily related to an orthogonal set.
\noindent

Of importance to us are OE sets that are up to a multiplicative
constant proportional to a unitary.
As we readily show, such orthogonal sets can be generated
by local operations and entanglement.
To see this, let us consider for simplicity a one-qubit POVM.
Suppose Alice wishes to apply $U_{AB}$ that leads to
the orthogonal Kraus set
$M_\mu =\alpha_\mu \sigma_\mu$.
To do that, Alice and Bob need the entangled state
\beq
|\Psi\ra_{Ab} = \sum_\mu \alpha_\mu |\mu\ra_A|\mu\ra_b .
\eeq
Bob starts by performing a local unitary transformation between his system
and his part ($b$) of the entangled state
\beq
U_{bB} = \sum_\mu |\mu\ra_{bb}\la\mu|\sigma_\mu ,
\eeq
and measures $b$ in a complementary basis $|\eta\ra_b$
with equal probability to get $\eta$.
This leads to
\beq
\sum_\mu \pm \alpha_\mu |\mu\ra_A \sigma_\mu ,
\eeq
where the  $\pm$ signs in front of each term is determined
according to Bob's outcomes for $\eta_b$.
Alice then can correct them all to be $+$,
according to the 2-bit message she received from Bob,
by applying an appropriate rotation on the ancilla.
Now we recall that a unitary acting on $A$ induces a unitary acting
on $M_\mu$.
Therefore, any Kraus set that is OE to the above
orthogonal $M_\mu$ can be generated
by Alice by means of an appropriate local unitary
followed by a measurement which records to result of the POVM.

More generally, in order to find what are the POVMs that can
be applied remotely, we need to check which Kraus sets are OE.

%\vspace{.1cm}
\noindent
{\bf Theorem}:
{\em Any n-qubits POVM can be represented by an OE Kraus operator set.}
%\vspace{.1cm}

Let us start with a one qubit POVM. Since $\sigma_\mu$ forms a
basis, we can expand
$M_\mu = \sum c_{\mu\eta}\sigma_\eta$,
where  $c_{\mu\eta}=(\sigma_\eta,M_\mu)$.
{\bf Lemma:} $M_\mu$ is OE iff $c^\dagger c$ is diagonal.
Proof: If $c^\dagger c$ is diagonal, the columns  of the matrix
$c$ are orthogonal vectors.
Hence $c$ may be expressed as a product of a unitary and
a diagonal matrix:
$c_{\mu\nu}= \sum_\eta U_{\mu\eta}\delta_{\eta\nu}\alpha_\nu$.
The reverse direction of the lemma is immediate.

Consider the conditions on the matrix $c$. From $\sum M_\mu^\dagger M_\mu=1$
we obtain   $\sum_{\mu\eta}|c_{\mu\eta}|^2 =1$ and
\beqa
\Re \sum_{\mu} c_{\mu 0}^*c_{\mu k}  =0, \label{real}\\
\Im \sum_{\mu} c_{\mu n}^*c_{\mu m} =0,   \label{imaginary}
\eeqa
where $\Re$ and $\Im$ denote the real and imaginary parts respectively,
and roman letters only run over $1,2,3$.
These conditions are  not enough to force every $M_\mu$ to
be OE. A general Kraus sum representation does
not admit a unitary equivalent orthogonal representation.
(e.g. $M_0\propto|\up_z\ra\la \up_x|$, $M_1\propto |\up_x\ra\la \down_x|$).

Consider however a general one-qubit POVM $F_\mu$.
Since $F_\mu$ are a semi-positive hermitian operators, they
 can be described by (at most) four hermitian Kraus operators
 $M_\mu = \sqrt{F_\mu}$ and consequently the matrix $c$ is real.
 Eq. (\ref{real}) than implies
that the 0'th and  $k$'th columns of $c$ are orthogonal.
Next suppose that after applying $U_{AB}$ in (\ref{unitary}),
Bob applies a local $\sigma_1$ rotation. Hence
$U_{AB}\to \sigma_1 U_{AB}$. This induces another Kraus set
obtained by
$c_{\mu 0} \to c_{\mu 1}$,
$c_{\mu 1} \to c_{\mu 0}$,
$c_{\mu 2} \to i c_{\mu 3}$,
$c_{\mu 3} \to -ic_{\mu 2}$.
But now, because columns 2 and 3 are purely imaginary,
we deduce from  eq. (\ref{imaginary}) that
the 1'st column must be orthogonal to the second and third columns.
Similarly we obtain that all columns are
orthogonal. Hence $\sqrt{F_\mu}$ is OE.
It is straightforward to generalize the above considerations to an n-qubit
POVM. This then concludes the proof.

As a corollary we conclude that:
{\em Any n-qubit POVM can by generated remotely by the present
method.}

Next let us quantify the entanglement resources
needed to apply a POVM.
The coefficients $\alpha_\mu$ fix the schmidt coefficients
of the needed shared entangled state.
This is readily found be noticing that $c^\dagger c$ given by a
diagonal matrix of the form
$\alpha_\mu^2\delta_{\mu\lambda}$. Therefore,
\beq
\label{alpha}
(\alpha_\lambda)^2 =
\sum_\mu (\sigma_\lambda, \sqrt{F_\mu})^2 .
\eeq
The above expression can be extended to n-qubit POVM
by replacing $\sigma_\mu$ with the basis
$\sigma^1_\mu\sigma^2_\lambda \cdots \sigma^n_\eta$.
The entanglement consumed for generating all  POVMs which
are unitary related to $\sqrt{F_\mu}$
is therefore
\beq
\label{measure}
E_{POVM} = -\sum_\lambda  \alpha^2_\lambda \ln \alpha^2_\lambda .
\eeq
The classical communication  cost is determined by the number $n$
of qubits on which we apply the POVM. It is at most given by $n$
bits.

We remark that in this approach Alice has full control on
the POVM and obtains the result of the measurement.
For other purposes it may be useful to "share" between
the job of performing the POVM between Alice and Bob.
The first example we gave (Fig. 1)
is indeed of that type. The entanglement needed for applying a
"shared" POVM is obviously smaller.

To summarize: we showed that any n-qubit POVM (that may be also viewed as a
set of generally non-unitary operations) can be
implemented remotely. To this end, for each class of unitary related
OE POVMs Alice and Bob need a particular  entangled state which is determined
by the POVM.
For the special case of a remote projective (von-Neumann) measurement
of $n$ qubits we have $E_{POVM}=n$.
After coupling the entangled particle to his system
Bob performs a measurement and transmits the result to Alice. She transforms
this state to the standard form of an orthogonal POVM.
To apply the POVM she applies the unitary
\beq
U_{\mu\nu} =   \sum_\eta (\sigma_\eta,  M_\nu)
\delta_{\eta\mu}{1\over\alpha_\eta}
\eeq
on her entangled ancilla and finally measures the
ancilla. The efficiency of the process
is optimal: i.e. the information gained is identical to a locally
performed POVM.

Is the measure given in (\ref{alpha},\ref{measure}) unique?.
Does it determine the minimal entanglement needed to
perform the POVM?
Let us show that if the action of the POVM on the system is
given as in our case, by the semi-positive hermitian
root $\sqrt{F_\mu}$, that is indeed true.

Consider the entanglement
capability of the remote POVM defined by
 $E_{capability} = \max_{\psi_B} E(\Psi_{AB})$.
 I.e. we maximize the entanglement generated by the POVM between
 the ancilla of Alice and Bob's system.
Next, let $E_{cost}$ be the minimal
entanglement needed to generate remotely the POVM.
By the  principle of  entanglement non-increase under LOCC,
 we must have
\beq \label{ineq1}
E_{capability} \le E_{cost}.
\eeq
We have seen that the POVM can be performed using the entanglement
$E_{POVM}$ defined in (\ref{measure}). Since our method may not be
optimal  it can be that $E_{cost} \le E_{POVM}$.
Now consider the entanglement capability. Since our POVM is OE
it can be transformed locally to
the form
$\sum|\mu\ra_A \alpha_\mu \sigma_\mu |\psi\ra_B$. Suppose that
Bob's particle entangled with another local particle
in an EPR state.
In this special case the entanglement capability
of the POVM is precisely $E_{POVM}$.
Therefore we arrive to the inequality
\beq
E_{capability} \ge E_{cost},
\eeq
which when combined with  (\ref{ineq1})
leads to the desired conclusion
\beq
E_{cost}=E_{capability}=E_{POVM}.
\eeq
Our  POVM construction therefore leads
to a unique one-to-one relations between POVMs and entanglement.
The entanglement $E_{POVM}$ constitutes a lower bound
on the entanglement cost\cite{remark} and an upper bound on the entanglement
capability.

I would like to thank Berry Groisman and Sandu Popescu for
very helpful discussions.
The research was supported in part by grant 62/01-1 of
the Israel Science Foundation, established by the
Israel Academy of Sciences and Humanities and by the Israel MOD
 Research and Technology Unit.


\begin{thebibliography}{99}
\bibitem{remark}
A POVM does not uniquely defined the back-action
on the observed system.
When the back-action can be described by a hermitian $\sqrt{F_\mu}$, up to 
additional local actions of Bob, our result is valide.
However consider unitaries performed remotetly by Alice
before the POVM. This requires
additional entanglement. It was not shown that in this case, 
the totat consumed entanglement is aways larger than $E_{POVM}$.
I thank S. Popescu for this remark.
\end{thebibliography}
\end{document}